\documentclass[10pt]{iopart}
\usepackage{iopams,setstack,epsfig,psfrag}
\usepackage{color}
\usepackage[svgnames]{xcolor}
\usepackage{rawfonts}
\input prepictex
\input pictex
\input postpictex
\def\@begintheorem#1#2{\par\bgroup{\sc #1\ #2. }\it\ignorespaces}
\def\@opargbegintheorem#1#2#3{\par\bgroup{\sc #1\ #2\ (#3). }\it\ignorespaces}
\def\@endtheorem{\egroup\par}

\eqnobysec

\usepackage{graphicx}
\usepackage{subfigure}
\usepackage{epstopdf}

\def\Pr{{\it Proof:  }}
\def\qed{$\Box$}
\def\L{\left(}   \def\R{\right)}
\def\LC{\left\{}   \def\RC{\right\}}
\def\q{\quad}
\def\qq{\qquad}

\def\lfl{\left\lfloor}
\def\rfl{\right\rfloor}

\def\C#1{{\cal{#1}}}

\def\Ref#1{(\ref{#1})}

\def\sfrac#1#2{\hbox{\normalsize $\frac{#1}{#2}$}}

\newtheorem{theo}{Theorem}
\newtheorem{lemm}{Lemma}
\newtheorem{corr}{Corollary}

\begin{document}

\title{Adsorbed self-avoiding walks subject to a force}
\author{E. J. Janse van Rensburg$^{1}$ and S. G. Whittington$^{2}$\\
$^1$Department of Mathematics and Statistics, York University, Toronto, Canada \\
$^2$Department of Chemistry, University of Toronto, Toronto, Canada
\\}

\begin{abstract}
We consider a self-avoiding walk model of polymer adsorption where the 
adsorbed polymer can be desorbed by the application of a force.  In this paper 
the force is applied normal to the surface at the last vertex of the walk.
We prove that the appropriate limiting free energy exists where there is  
an applied force and a surface potential term, and prove that this free energy
is convex in appropriate variables.  We then derive an expression for the 
limiting free energy in terms of the free energy without a force and the free energy with no surface 
interaction.  Finally we show that there is a phase boundary between the 
adsorbed phase and the desorbed phase in the presence of a force, prove some qualitative 
properties of this boundary and derive bounds on the location of the boundary.

\end{abstract}

\maketitle

\section{Introduction}
\setcounter{equation}{0}

Polymer adsorption at an impenetrable surface \cite{DeBell}
is a topic that has been attracting interest
from theorists for at least fifty years \cite{Rubin,Silberberg1962}.  Various polymer models have
been examined, including random walks \cite{Rubin,Binder2012}, directed and partially directed 
walks \cite{Forgacs,Rensburg2003,Whittington1998} and self-avoiding walks \cite{HTW1982,Hegger1994,Rensburg1998,Rensburg2004}.
Techniques like atomic force microscopy \cite{Haupt1999} allow adsorbed polymers to be pulled 
off a surface and this problem has also attracted attention \cite{Krawczyk2005,Orlandini1999,Owczarek2010,Binder2012}.  
Variants of the problem such as pulling at an angle 
\cite{Orlandini2010,Osborn2010,Serr2007,Tabbara} and from inhomogeneous surfaces \cite{Iliev2013,Iliev2012}
have also been studied.

The standard lattice model of polymer adsorption is self-avoiding walks.  In the absence of a 
force this problem is quite well understood \cite{DeBell,HTW1982,Rensburg1998}, though many details remain as open questions.  For instance, although there is a rigorous proof of the 
existence of a phase transition, the order of the transition is not known rigorously.  When the 
adsorbed self-avoiding walk is subject to a force, numerical evidence shows that self-avoiding
walks are well approximated by partially directed walks \cite{Iliev2013}, at least for 
large forces.  Nevertheless, at the rigorous level we know very little about  the behaviour of 
self-avoiding walks in these circumstances.  The phase transition (\emph{i.e.} desorption)
under the influence of a 
force is expected to be first order but this is not known rigorously.

%%%%%%%%%%%%%%%%%%%%%%%%%%%%%%%%%%%%%%%%%%%%%%%%%%
\begin{figure}[h!]
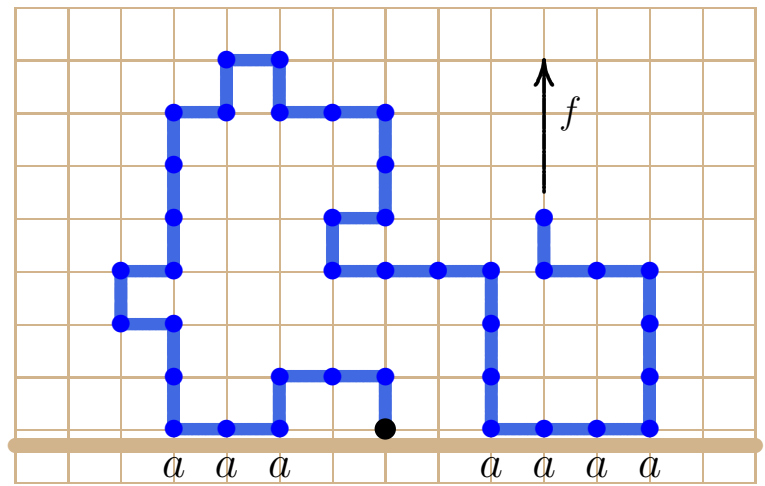

\beginpicture
\setcoordinatesystem units <2pt,2pt>
\setplotarea x from 0 to 140, y from -10 to 80
\color{Tan}
\put {\beginpicture \grid 14 9 \endpicture} at 40 0

\setcoordinatesystem units <2pt,2pt> point at -40 0 
\color{Tan}
\setplotsymbol ({\Large$\bullet$})
\plot 0 -3 140 -3 /

\setplotsymbol ({\large$\bullet$})
\color{RoyalBlue}
\plot 70 0 70 10 60 10 50 10 50 0 40 0 30 0 30 10
30 20 20 20 20 30 30 30 30 40 30 50 30 60 40 60 40 70 
50 70 50 60 60 60 70 60 70 50 70 40 60 40 60 30 70 30
80 30 90 30 90 20 90 10 90 0 100 0 110 0 120 0 120 10 
120 20 120 30 110 30 100 30 100 40 /

\color{black}
\setplotsymbol ({\large$\cdot$})
\arrow <10pt> [.2,.67] from 100 45 to 100 70
\put {\Large$f$} at 105 60
\multiput {\LARGE$a$} at 30 -7 40 -7 50 -7 90 -7 
100 -7 110 -7 120 -7 /

\color{blue}
\multiput {{\LARGE$\bullet$}} at 
70 0 70 10 60 10 50 10 50 0 40 0 30 0 30 10
30 20 20 20 20 30 30 30 30 40 30 50 30 60 40 60 40 70 
50 70 50 60 60 60 70 60 70 50 70 40 60 40 60 30 70 30
80 30 90 30 90 20 90 10 90 0 100 0 110 0 120 0 120 10 
120 20 120 30 110 30 100 30 100 40 /

\color{black}
\put {\huge$\bullet$} at 70 0 

\normalcolor
\endpicture
\caption{A pulled and adsorbing self-avoiding walk in the positive
half-lattice.  Visits of the walk to the adsorbing line are
weighted by $a$, while a force $f$ is pulling the path at its
endpoint in the vertical direction.}
\label{figure1}   %%ZXZ[figure1]
\end{figure}
%%%%%%%%%%%%%%%%%%%%%%%%%%%%%%%%%%%%%%%%%%%%%%%%%%

In this paper we consider the simplest case, that of a self-avoiding walk terminally attached
to an impenetrable surface at which the walk adsorbs, and pulled from the other 
unit degree vertex  in a direction \emph{normal} to the surface.  We prove the existence of the 
appropriate thermodynamic limit and some qualitative properties of the limiting
free energy.  In addition we derive some information about the form of the phase 
diagram for the model.

\section{Thermodynamic limits and convexity}
\setcounter{equation}{0}
\label{sec:thermolim}

Consider the $d$-dimensional hypercubic lattice $Z^d$ where the vertices have 
integer coordinates $(x,y,\ldots z)$.  We write $(x_i,y_i, \ldots z_i)$, 
$i=0,1,2, \ldots n$ for the coordinates of the $i$-th vertex of an 
$n$-step self-avoiding walk on $Z^d$. 
The number of $n$-step self-avoiding walks  from the origin is denoted by
$c_n$. It is known that $\lim_{n\to\infty} \sfrac{1}{n} \log c_n 
= \log \mu_d$ exists \cite{HM54}, where $\mu_d$ is the 
\emph{growth constant} of self-avoiding walks.

A \emph{positive walk} is a self-avoiding walk on $Z^d$ that starts 
at the origin and is constrained to have $z_i \ge 0$ for all 
$0 \le i \le n$.  The number of $n$-step positive walks from 
the origin is denoted by $c_n^+$.  It is known that 
$\lim_{n\to\infty} \sfrac{1}{n} \log c_n^+ = \log \mu_d$ \cite{HTW1982}.
Vertices of a positive walk in the hyperplane $z=0$ are 
\emph{visits} although, by convention, the vertex at the origin 
is not counted as a visit. The number 
of positive walks of $n$-steps from the 
origin with $v$ visits is denoted by $c_n^+(v)$.

An $n$-step  positive walk is a \emph{loop} if $z_n = 0$.  
A positive walk (or a loop) is \emph{unfolded in the $x$-direction} 
if $0 \le x_i < x_n$, $i=1,2, \ldots n-1$, with similar definitions for the 
$y$-direction, up to and including the $z$-direction. See \cite{HammersleyWelsh}.

%%%%%%%%%%%%%%%%%%%%%%%%%%%%%%%%%%%%%%%%%%%%%%%%%%
\begin{figure}[h!]
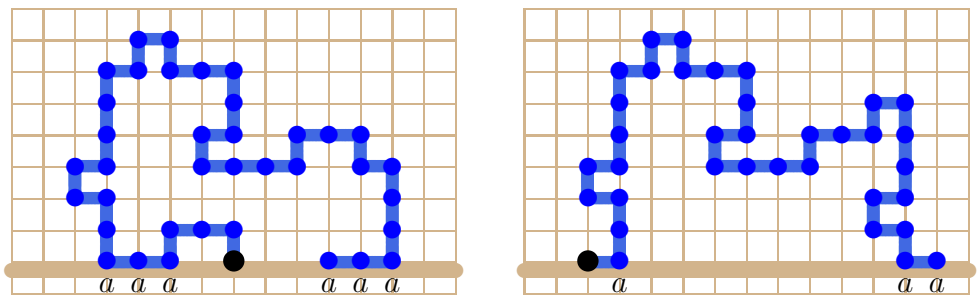

\beginpicture

\put 
{\beginpicture
\setcoordinatesystem units <1.2pt,1.2pt>
\setplotarea x from 0 to 140, y from -10 to 80
\color{Tan}
\put {\beginpicture \grid 14 9 \endpicture} at 40 0

\setcoordinatesystem units <1.2pt,1.2pt> point at -40 0 
\color{Tan}
\setplotsymbol ({\Large$\bullet$})
\plot 0 -3 140 -3 /

\color{RoyalBlue}
\setplotsymbol ({\large$\bullet$})
\plot 70 0 70 10 60 10 50 10 50 0 40 0 30 0 30 10
30 20 20 20 20 30 30 30 30 40 30 50 30 60 40 60 40 70 
50 70 50 60 60 60 70 60 70 50 70 40 60 40 60 30 70 30
80 30 90 30 90 40 100 40 110 40 110 30 120 30 120 20 
120 10 120 0 110 0 100 0 /

\color{black}
\setplotsymbol ({\large$\cdot$})
%\arrow <10pt> [.2,.67] from 100 45 to 100 70
%\put {\Large$f$} at 105 60
\multiput {\large$a$} at 30 -7 40 -7 50 -7  
100 -7 110 -7 120 -7 /

\color{blue}
\multiput {{\LARGE$\bullet$}} at 
 70 0 70 10 60 10 50 10 50 0 40 0 30 0 30 10
30 20 20 20 20 30 30 30 30 40 30 50 30 60 40 60 40 70 
50 70 50 60 60 60 70 60 70 50 70 40 60 40 60 30 70 30
80 30 90 30 90 40 100 40 110 40 110 30 120 30 120 20 
120 10 120 0 110 0 100 0 /

\color{black}
\multiput {\huge$\bullet$} at 70 0  /

\normalcolor
\endpicture} at 0 0 

\put
{\beginpicture
\setcoordinatesystem units <1.2pt,1.2pt> point at 0 0 
\setplotarea x from 0 to 140, y from -10 to 80
\color{Tan}
\put {\beginpicture \grid 14 9 \endpicture} at 40 0

\setcoordinatesystem units <1.2pt,1.2pt> point at -40 0 
\color{Tan}
\setplotsymbol ({\Large$\bullet$})
\plot 0 -3 140 -3 /

\color{RoyalBlue}
\setplotsymbol ({\large$\bullet$})
\plot 20 0 30 0 30 10 30 20 20 20 20 30 30 30 30 40 
30 50 30 60 40 60 40 70 50 70 50 60 60 60 70 60 70 50 
70 40 60 40 60 30 70 30 80 30 90 30 90 40 100 40 110 40
110 50 120 50 120 40 120 30 120 20 110 20 110 10 120 10
120 0 130 0 /

\color{black}
\setplotsymbol ({\large$\cdot$})
%\arrow <10pt> [.2,.67] from 100 45 to 100 70
%\put {\Large$f$} at 105 60
\multiput {\large$a$} at 30 -7 120 -7 130 -7 /

\color{blue}
\multiput {{\LARGE$\bullet$}} at 
20 0 30 0 30 10 30 20 20 20 20 30 30 30 30 40 
30 50 30 60 40 60 40 70 50 70 50 60 60 60 70 60 70 50 
70 40 60 40 60 30 70 30 80 30 90 30 90 40 100 40 110 40
110 50 120 50 120 40 120 30 120 20 110 20 110 10 120 10
120 0 130 0 /

\color{black}
\put {\huge$\bullet$} at 20 0 

\normalcolor
\endpicture} at 200 0

\endpicture
\caption{A loop and an unfolded loop on the square lattice.}
\label{figure2}   %%ZXZ[figure2]
\end{figure}
%%%%%%%%%%%%%%%%%%%%%%%%%%%%%%%%%%%%%%%%%%%%%%%%%%

The number of $n$-step loops with $v$ visits is denoted $l_n(v)$.
Similarly we write $c^{\ddagger}_n(v)$ and $l^{\ddagger}_n(v)$ 
for the corresponding numbers of positive walks and loops that 
are unfolded in the $x$-direction.  
We define the corresponding partition functions
\begin{equation}
C_n(a) = \sum_v c_n^+(v)\, a^v, \qq 
L_n(a) = \sum_v l_n(v)\, a^v
\label{eqn2-1}    %%ZXZ[eqn2-1]
\end{equation}
and 
\begin{equation}
C_n^{\ddagger}(a) = \sum_v c^{\ddagger}_n(v)\, a^v,\qq
L^{\ddagger}_n(a) = \sum_v l^{\ddagger}_n(v)\, a^v.
\end{equation}
A \emph{tail} is a positive walk that leaves the surface on its 
first step and never returns.  Clearly the number of $n$-step 
tails is $c_n^+(0)$.  

Next we recall a useful lemma.
\begin{lemm}
For unfolding in the $x$-direction and $a>0$ the following limits exist and are equal:
$$
\lim_{n\to\infty} \sfrac{1}{n} \log C_n(a) 
= \lim_{n\to\infty} \sfrac{1}{n}  \log L_n(a) 
= \lim_{n\to\infty} \sfrac{1}{n} \log C^{\ddagger}_n(a) 
= \lim_{n\to\infty} \sfrac{1}{n}  \log L^{\ddagger}_n(a) \equiv \kappa(a).
$$
Also, $\kappa(a) = \log \mu_d$ for $a \leq 1$. \qed
\label{lem:loops}
\end{lemm}
See \cite{HTW1982} for the proof of this lemma.  

We generalize the model above to a two-parameter model of \emph{pulled
adsorbing positive walks}.

Let $c_n^+(v,h)$ be the number of $n$-step positive walks with 
$v$ visits and with $z_n=h$.  Then $c_n^+(0,h)$ is the number of 
$n$-step tails with $z_n=h$. We call $h$ the \emph{height}
of the last vertex.

We write $c^{\ddagger}_n(v,h)$ for the numbers of the corresponding 
unfolded walks.  The corresponding partition functions are
\begin{equation}
C_n(a,y) = \sum_{v,h} c_n^+(v,h)\, a^v y^h, \qq
T_n(y) = \sum_h c_n^+(0,h)\, y^h
\label{equation2.3}
\end{equation}
and 
\begin{equation}
C_n^{\ddagger}(a,y) = \sum_{v,h} c^{\ddagger}_n(v,h)\, a^v y^h,\qq
T^{\ddagger}_n(y) = \sum_h c^{\ddagger}_n(0,h)\, y^h.
\end{equation}

The following lemma was proved in \cite{Rensburg2009}.
\begin{lemm}
For unfolding in the $x$-direction the following limits exist and 
are equal:
$$\lim_{n\to\infty} \sfrac{1}{n}  \log T_n(y) 
= \lim_{n\to\infty} \sfrac{1}{n}  \log T^{\ddagger}_n(y) 
= \lambda (y).$$
\label{lem:tails}
\end{lemm}

%%%%%%%%%%%%%%%%%%%%%%%%%%%%%%%%%%%%%%%%%%%%%%%%%%
\begin{figure}[h!]
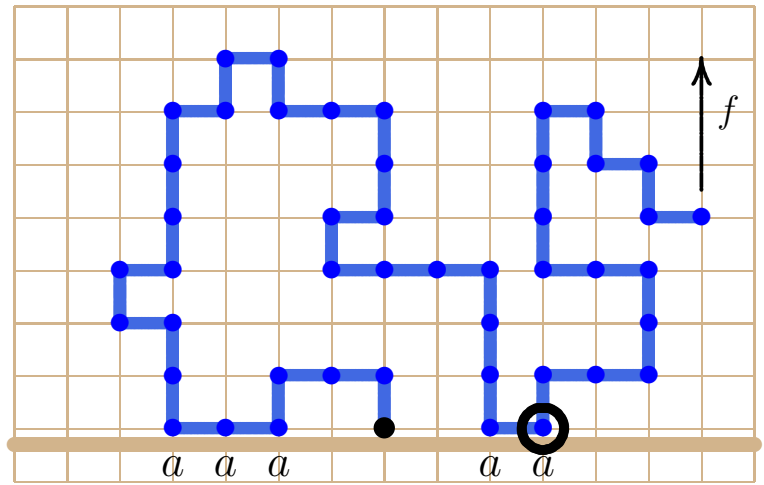

\beginpicture
\setcoordinatesystem units <2pt,2pt>
\setplotarea x from 0 to 140, y from -10 to 80
\color{Tan}
\put {\beginpicture \grid 14 9 \endpicture} at 40 0

\setcoordinatesystem units <2pt,2pt> point at -40 0 
\color{Tan}
\setplotsymbol ({\Large$\bullet$})
\plot 0 -3 140 -3 /

\color{RoyalBlue}
\setplotsymbol ({\large$\bullet$})
\plot 70 0 70 10 60 10 50 10 50 0 40 0 30 0 30 10
30 20 20 20 20 30 30 30 30 40 30 50 30 60 40 60 40 70 
50 70 50 60 60 60 70 60 70 50 70 40 60 40 60 30 70 30
80 30 90 30 90 20 90 10 90 0 100 0 100 10 110 10 120 10 
120 20 120 30 110 30 100 30 100 40 100 50 100 60 110 60
110 50 120 50 120 40 130 40  /

\color{black}
\setplotsymbol ({\large$\cdot$})
\arrow <10pt> [.2,.67] from 130 45 to 130 70
\put {\Large$f$} at 135 60
\multiput {\LARGE$a$} at 30 -7 40 -7 50 -7 90 -7 100 -7 /

\setplotsymbol ({\small$\bullet$})
\circulararc 360 degrees from 104 0 center at 100 0

\color{blue}
\multiput {{\LARGE$\bullet$}} at 
70 0 70 10 60 10 50 10 50 0 40 0 30 0 30 10
30 20 20 20 20 30 30 30 30 40 30 50 30 60 40 60 40 70 
50 70 50 60 60 60 70 60 70 50 70 40 60 40 60 30 70 30
80 30 90 30 90 20 90 10 90 0 100 0 100 10 110 10 120 10 
120 20 120 30 110 30 100 30 100 40 100 50 100 60 110 60
110 50 120 50 120 40 130 40 /

\color{black}
\put {\huge$\bullet$} at 70 0 

\normalcolor
\endpicture
\caption{The circled vertex is the last visit of the walk.  At this vertex the walk
can be decomposed into a loop and a tail.}
\label{figure3}   %%ZXZ[figure3]
\end{figure}
%%%%%%%%%%%%%%%%%%%%%%%%%%%%%%%%%%%%%%%%%%%%%%%%%%

We extend these two lemmas by proving the existence of the thermodynamic 
limit for the two parameter adsorbing and pulled positive walk model.
\begin{theo}
The limit 
$\lim_{n\to\infty} \sfrac{1}{n} \log C_n(a,y)$
exists and is equal to $\max[\kappa (a), \lambda (y)]$.
\label{theo:thermolim}
\end{theo}
\Pr
Every positive walk is either a loop or it eventually leaves the 
surface for the last time.  This decomposes positive walks into
a loop, followed by a tail (either the loop, or the tail, may be
empty) -- see Figure \ref{figure3}.

If the loop and the tail in the decomposition are both unfolded,
then a lower bound is obtained:  
$$C_n(a,y) \ge \sum_{m=0}^n L^{\ddagger}_m(a) \, T^{\ddagger}_{n-m}(y)$$
since the two unfolded parts must be distinct except for the vertex where the 
loop ends and the tail begins.  
In particular, by only considering the $m=0$ or $m=n$ terms on
the right hand side, the following lower bound on $C_n(a,y)$ is obtained:
$$C_n(a,y) \ge \max[L^{\ddagger}_n(a),T^{\ddagger}_n(y)]$$
and therefore 
$$\liminf_{n\to\infty} \sfrac{1}{n} \log C_n(a,y) \ge \max[\kappa(a),\lambda(y)]$$
where we have made use of Lemma \ref{lem:loops} and Lemma \ref{lem:tails}.

To obtain an upper bound on $C_n(a,y)$ we again consider the loop plus
tail decomposition in Figure \ref{figure3}.  If all loops and tails are
considered, then an upper bound on $C_n(a,y)$ follows:
$$C_n(a,y) \le \sum_m L_m(a)\, T_{n-m}(y).$$
Define the generating functions
$$\widehat{L}(a,t) = \sum_n L_n(a)\, t^n, \qq 
\widehat{T}(y,t) = \sum_n T_n(y)\, t^n.$$
Using the convolution theorem 
$$\widehat{L}(a,t)\, \widehat{T}(y,t) 
= \sum_n\sum_{m=0}^n L_m(a)\, T_{n-m}(y)\, t^n 
= \sum_n B_n(a,y)\, t^n = \widehat{B}(a,y,t)$$
and so $C_n(a,y) \le B_n(a,y)$.  

The radius of convergence of 
$\widehat{L}(a,t)$ is $t=t_1=\exp(-\kappa(a))$ and of
$\widehat{T}(y,t)$ is $t=t_2=\exp(-\lambda(y))$.  Thus, 
the product $\widehat{B}(a,y,t)$ has radius of convergence
$\min\{t_1,t_2\}$ and is singular at $t= t_1$ and at $t=t_2$.  

Hence $\limsup_{n\to\infty} \sfrac{1}{n} \log B_n(a,y) = \max[\kappa(a),
\lambda(y)]$ and therefore 
$$\limsup_{n\to\infty} \sfrac{1}{n} \log C_n(a,y) 
\le \max[\kappa(a),\lambda(y)].$$
Hence the thermodynamic limit exists and 
$\lim_{n\to\infty} \sfrac{1}{n} \log C_n(a,y) = \max[\kappa(a),\lambda(y)].$
\qed

We shall write 
\begin{equation}
\lim_{n\to\infty} \sfrac{1}{n} \log C_n(a,y) =\kappa(a,y)
\label{FE}
\end{equation}
so $\kappa(a,y) = \max[\kappa(a),\lambda(y)]$.

\begin{theo}
The limiting  free energy 
$\kappa(a,y)=\lim_{n\to\infty} \sfrac{1}{n} \log C_n(a,y)$ is a 
convex function of $\log a$ and $\log y$.
\label{convexity}
\end{theo}
\Pr
The theorem is a consequence of H\"{o}lder's inequality.  
For any $p$ such that $0 \le p \le 1$, 
\begin{equation}
\hspace{-1cm}
\sum_{v,h} c_n^+(v,h)\, (a_1^{p }a_2^{1-p})^v 
\,(y_1^{p} y_2^{1-p})^h \le
\left(\sum_{v,h} c_n^+(v,h)\,a_1^vy_1^h\right)^{p} 
\left(\sum_{v,h} c_n^+(v,h)\,a_2^vy_2^h\right)^{1-p}
\end{equation}
Taking logarithms and dividing by $n$ gives
\begin{equation}
\sfrac{1}{n} \log C_n(a_1^{p}a_2^{1-p}, y_1^p y_2^{1-p}) 
\le p \, \sfrac{1}{n} \log C_n(a_1,y_1) + (1-p)\, 
\sfrac{1}{n} \log C_n(a_2,y_2)
\end{equation}
so $\sfrac{1}{n} \log C_n(a,y)$ is a convex function of $\log a$ 
and $\log y$.  Since the limit of a sequence of convex functions, 
when it exists, is also convex, this proves the theorem.
 \qed

It follows that the free energy is convex as a surface and not only in the coordinate directions.

\section{The free energy}

In this section we turn our attention to the free energy $\kappa(a,y)$ 
given by equation (\ref{FE}) in terms of $\kappa(a)$ 
(the free energy of an adsorbing walk) and $\lambda(y)$ (the free
energy of a pulled walk).  We first review the properties of 
$\kappa(a)$ and $\lambda(y)$, and prove that $\kappa (a)$ is asymptotic
to $\log\mu_{d-1} + \log a$ and $\lambda(y)$ is asymptotic to $\log y$.

\subsection{Bounds on $\kappa (a)$}

We first recall some earlier results for the adsorption problem without 
a force \cite{HTW1982}.

It is known that $\kappa (a)$ is singular at a critical point
at $a=a_c^o$ corresponding to the adsorption transition in this model.
It is known that $1 < a_c^o < \sfrac{\mu_d}{\mu_{d-1}}$, 
see for example \cite{HTW1982}, or  \cite{Rensburg2000B} for a review.  
The order of the adsorption transition is not known rigorously
but is thought to be continuous \cite{HTW1982,Hegger1994,Rensburg2004}. For $a < a_c^o$, 
$\kappa (a) = \kappa (1) = \log \mu_d$, independent of $a$, 
and for $ a > a_c^o$
\begin{equation}
\max \LC \log \mu_d,\, \log \mu_{d-1} + \log a \RC
\le \kappa (a) 
\le \log \mu_d + \log a.
\label{eqn3.1}  %%ZXZ[eqn3.1]
\end{equation}
The average number of visits in an adsorbing loop is given by
\begin{equation}
\langle v \rangle_n = a\sfrac{d}{da} \log \sum_v l_n(v) \,a^v = 
\frac{ \sum_v v\, l_n(v) \,a^v }{\sum_v l_n(v) \,a^v} ;
\label{eqn3.2}  %%ZXZ[eqn3.2]
\end{equation}
see equation \Ref{eqn2-1}.
Since $\sfrac{1}{n} \log L_n(a)$ is a sequence of convex 
functions converging to a convex limit $\kappa (a)$, 
one may interchange the limit and the derivative 
almost everywhere in
\begin{equation}
{\cal E}(a) = \lim_{n\to\infty} \sfrac{1}{n} \langle v \rangle_n 
= \lim_{n\to\infty} \sfrac{a}{n} \sfrac{d}{da} \log L_n(a)
= a\,\sfrac{d}{da} \kappa (a)
\label{eqn3.3}  %%ZXZ[eqn3.3]
\end{equation}
where ${\cal E} (a)$ is equal to the density of visits almost everywhere.

The left- and right-densities of visits may be defined as follows
\begin{equation}
{\cal E}_-(a) = a\,\sfrac{d^-}{da} \kappa (a),\qq
{\cal E}_+(a) = a\,\sfrac{d^+}{da} \kappa (a).
\end{equation}
These exist for every finite $a>0$ since $\kappa(a)$
is a convex function of $\log a$.  Moreover ${\cal E}_-(a)$ is 
lower semi-continuous and ${\cal E}_+(a)$ is upper semi-continuous.
It follows that ${\cal E}_- (a) = {\cal E}_+ (a) = {\cal E}(a)$
for almost every $a$, and ${\cal E}_- (a) \leq {\cal E}_+(a)$
for every $a>0$.

The density of visits, and ${\cal E}_-(a)$ and ${\cal E}_+(a)$
are monotone functions, and they are differentiable almost 
everywhere. Clearly ${\cal E} (a) = 0$ if $a < a_c^o$ 
and ${\cal E}(a) > 0$ for almost all $a>a_c^o$.

The Legendre transform of $\kappa(a)$ is defined by
\begin{equation}
\log {\cal P} (\alpha) 
= \inf_{a>0} \LC \kappa(a) - \alpha\,\log a \RC .
\end{equation}
The function $\log {\cal P} (\alpha)$ is a concave function 
of $\alpha$ (see for example \cite{Rensburg2000B}).  
The inverse transform is
\begin{equation}
\kappa (a) = \sup_{\alpha\in[0,1]} \left\{ \log {\cal P}(\alpha) + \alpha \log a \right\},
\label{eqn3.5}
\end{equation}
see reference \cite{Rensburg2000B}.

The supremum in equation \Ref{eqn3.5} may be realised when $\alpha=\alpha^*=0$ or
$\alpha=\alpha^*=1$, in which case $\kappa(a) = \log {\cal P} (0)$ or
$\kappa(a) = \log {\cal P} (1) + \log a$. (This occurs at boundaries of the
range of $\alpha$ in equation \Ref{eqn3.5}). In these cases we note by
equation \Ref{eqn3.3} that ${\cal E}(a) = 0 = \alpha^*$ or
${\cal E}(a) = 1 = \alpha^*$, respectively.

Otherwise we notice that $\log {\cal P}(\alpha) + \alpha \log a$ is
a concave function of $\log a$.  Hence, it has left-derivatives and
right-derivatives everywhere.  Since $\kappa(a)$ is defined for every
$a\geq 0$, there exists an $\alpha^* \in [0,1]$ solving equation
\Ref{eqn3.5}.  If $\alpha^* \not= 0$ and $\alpha^* \not= 1$, then it
must be in $(0,1)$, in which case one may instead consider
the solutions of
\begin{equation}
\frac{d^-}{d\alpha} \left( \log {\cal P}(\alpha) + \alpha \log a \right) = 0,
\quad\hbox{or}\;\;
\frac{d^+}{d\alpha} \left( \log {\cal P}(\alpha) + \alpha \log a \right) = 0.
\end{equation} 
The solutions of these ($\alpha_-$ and $\alpha_+$, respectively)
will be equal for almost every $a$ (because of
concavity).  Notice that the solutions will be continuous almost
everywhere, except where $\log {\cal P}(\alpha) + \alpha \log a$ is
not differentiable.  In that case $\alpha_- < \alpha_+$ and since
$\log {\cal P}(\alpha)$ is a concave function of $\alpha$,
both $\alpha_-$ and $\alpha_+$ are monotone and so differentiable
almost everywhere, while $\alpha_- \leq \alpha^* \leq \alpha_+$.

In addition, $\kappa(a) = \log {\cal P}(\alpha_-) + 
\alpha_- \log a$ for almost all $a>0$, and similarly,
$\kappa(a) = \log {\cal P}(\alpha_+) + 
\alpha_+ \log a$ for almost all $a>0$.

Observe that $\sfrac{d^-}{d\alpha}\log {\cal P}(\alpha) 
= - \log a$ if $\alpha = \alpha_-$ and 
$\sfrac{d^+}{d\alpha}\log {\cal P}(\alpha) = - \log a$
if $\alpha = \alpha_+$.
 
Taking the left derivative of $\kappa(a)$ with respect to $\log a$ gives
the left-density of visits ${\cal E}_-(a)$ which exists 
everywhere.  Using the chain rule for differentiation it follows that
for almost every $a$,
\begin{eqnarray*}
{\cal E}_-(a) &= a\frac{d^-}{da} \kappa (a)
 = a\frac{d^-}{da} \L \log {\cal P}(\alpha_-) + \alpha_- \log a\R \cr
&= \left[a\frac{d^-}{da}  \alpha_-  \right]
   \,\frac{d^-}{d\alpha} 
    \log {\cal P}(\alpha)\vert_{\alpha = \alpha_-}
 + a\log a\, \left[\frac{d^-}{da} \alpha_-\right] + \alpha_- \cr
&=  \left[a\frac{d^-}{da} \alpha_-   \right]
   \,(- \log a)  + a\log a\, \left[\frac{d^-}{da} \alpha_-\right] + \alpha_- = \alpha_- .
\end{eqnarray*}
%This is true for all $a$ since $\log {\cal P}(\alpha)$ is a concave function 
%of $\alpha$, and has left- and right-derivatives everywhere.
This is true for almost all $a$ since $\log {\cal P}(\alpha)$ is a 
concave function of $\alpha$, and has left- and right-derivatives 
everywhere while $\alpha_-$ is monotone and has left-derivatives 
almost everywhere. 

Similarly, one may show that
${\cal E}_+(a) = a\frac{d^+}{da} \kappa (a) = \alpha_+$.
Since $\alpha_-=\alpha_+$ almost everywhere, it follows that
${\cal E}_+(a) = {\cal E}_-(a)$ almost everywhere and is
equal to $a\frac{d}{da} \kappa (a)$ almost everywhere.

Thus, ${\cal E}_+(a) = {\cal E}_-(a) = {\cal E}(a)$ almost everywhere,
and we may redefine ${\cal E}(a) = {\cal E}_-(a) = \alpha_- = \alpha^*$.
This shows that the supremum in equation \Ref{eqn3.5} is 
realized by $\alpha^*$, or by the density of visits, for almost
every $a$.  Hence
\begin{equation}
\lim_{n\to\infty} \sfrac{1}{n} \langle v \rangle_n = {\cal E}(a) = \alpha^*, \q\hbox{for almost every $a>0$}.
\end{equation}

Define the function $v^*_n = \lfl \langle v \rangle_n \rfl$.  
Then we have proven the following theorem.

\begin{theo}
For almost every value of $a>0$,
$$\lim_{n\to\infty} \sfrac{1}{n}\, v^*_n 
= {\cal E}(a) = \alpha^*$$
where $\alpha^*$ realises the supremum in equation
\Ref{eqn3.5}. \qed
\label{thm3-7}
\end{theo}

Observe that $v^*_n \to \infty$ as $n\to\infty$ if
$a>a_c^o$.
We proceed by finding a lower bound on the partition 
function $L_n(a)$ in terms of unfolded loops.

First consider walks from the origin and confined to the
adsorbing plane.  Suppose the walks are unfolded in the $x$-direction 
and are of length $v^*_n$. The number of such walks is 
denoted by $c_{v^*_n}^{(d-1)}$, since these are unfolded 
self-avoiding walks in the $(d{-}1)$-dimensional lattice
which is the adsorbing plane.  The number of visits of
these walks is $v^*_n$.

Next, consider unfolded loops of length $n-v^*_n$ with 
exactly $1$ visit (the endpoint).  Such loops start in the
origin, step into the positive half-lattice above the
adsorbing plane, and only return to the adsorbing plane
at their  final step.  The number of such loops is 
$l_{n-v^*_n}^\ddagger(1)$.  

%%%%%%%%%%%%%%%%%%%%%%%%%%%%%%%%%%%%%%%%%%%%%%%%%%
\begin{figure}[h!]
\beginpicture
\setcoordinatesystem units <2pt,2pt>
\setplotarea x from 0 to 140, y from -10 to 80
\color{Tan}
\put {\beginpicture \grid 14 9 \endpicture} at 40 0

\setcoordinatesystem units <2pt,2pt> point at -40 0 
\color{Tan}
\setplotsymbol ({\Large$\bullet$})
\plot 0 -3 140 -3 /

\color{RoyalBlue}
\setplotsymbol ({\large$\bullet$})
\plot 20 0 30 0 40 0 50 0 60 0 70 0 70 10 80 10 80 20 
80 30 70 30 70 40 70 50 80 50 90 50 100 50 100 60 110 60 
110 50 110 40 120 40 120 30 110 30 100 30 90 30 90 20 
90 10 100 10 110 10 120 10 130 10 130 0  /

\color{black}
\setplotsymbol ({\large$\cdot$})
\multiput {\LARGE$a$} at 30 -7 40 -7 50 -7 60 -7 70 -7
130 -7 /

\setplotsymbol ({\small$\bullet$})
\circulararc 360 degrees from 74 0 center at 70 0

\color{blue}
\multiput {{\LARGE$\bullet$}} at 
20 0 30 0 40 0 50 0 60 0 70 0 70 10 80 10 80 20 
80 30 70 30 70 40 70 50 80 50 90 50 100 50 100 60 110 60 
110 50 110 40 120 40 120 30 110 30 100 30 90 30 90 20 
90 10 100 10 110 10 120 10 130 10 130 0 /

\color{black}
\put {\huge$\bullet$} at 20 0 

\normalcolor
\endpicture
\caption{A loop which lies in the adsorbing plane, leaves this plane and returns for the 
first time at its last step.  The circled vertex is the penultimate visit.}
\label{figure4}   %%ZXZ[figure4]
\end{figure}
%%%%%%%%%%%%%%%%%%%%%%%%%%%%%%%%%%%%%%%%%%%%%%%%%%

Concatenate these unfolded loops of length $n-v^*_n$
with the unfolded walks of length $v^*_n$ from the origin
in the adsorbing plane (by placing the first vertex
of the unfolded loop on the last vertex of the unfolded walk
-- see Figure \ref{figure4}).  This shows that
\begin{equation}
c_{v^*_n}^{(d-1)} \, l_{n-v^*_n}(1) \, a^{v^*_n+1}
\leq L_n (a),
\label{eqn3.8A}  
\end{equation}
since each concatenated pair contributes a term to the
partition function of adsorbing loops (notice that there
are $v^*_n+1$ total visits in the concatenated pair).

Take the logarithms of equation \Ref{eqn3.8A}, divide by $n$
and let $n\to\infty$.  If $a>a_c^o$, then $v^*_n \to\infty$
and $\lim_{n\to\infty} \sfrac{1}{n} v^*_n = {\cal E}(a) = \alpha^*$
for almost every $a$ by theorem \ref{thm3-7}.  
By Lemma \ref{lem:loops} it follows that
\begin{equation}
\fl \hspace{1cm}
\left( 1- {\cal E}(a) \right)\,\log \mu_d 
+ {\cal E}(a)\,\left(\log\mu_{d-1} + \log a \right)
\leq \kappa(a),\quad\hbox{for almost every $a>a_c^o$} .
\label{eqn3.8}  %%%ZXZ[eqn3.8]
\end{equation}
Since ${\cal E}(a) = a\frac{d}{da} \kappa(a)$ for almost
every $a>0$, this gives the following lemma if $a=e^x$.

\begin{lemm}
Suppose that $f(x) = \kappa(e^x)$. Then for almost every 
$x > \log a_c^o$,
\[ f(x) - \log \mu_d 
\geq  \left(x - \log \frac{\mu_d}{\mu_{d-1}}\right) 
\, \frac{d}{dx} f(x) . \]
\label{lemA}  %%ZXZ[lemA]
\qed
\end{lemm}

This inequality can be integrated from $\log z$ to $\log a$
where $z > \sfrac{\mu_d}{\mu_{d-1}}$.
This gives the following theorem.

\begin{theo}
The free energy of adsorbing walks satisfies the
following inequality:
\[ \frac{\kappa(a) - \log \mu_d}{\log a -\log\frac{\mu_d}{\mu_{d-1}}}
\leq \frac{\kappa(z) - \log \mu_d}{\log z-\log\frac{\mu_d}{\mu_{d-1}}}\]
whenever $a > z> \frac{\mu_d}{\mu_{d-1}}$.
\label{thmA}  %%ZXZ[thmA]
\end{theo}
\Pr
Separate variables in Lemma \ref{lemA} and integrate 
over $(\log z,\log a)$ where $z>\frac{\mu_d}{\mu_{d-1}}$.  This gives
\[ \int_{\log z}^{\log a} \frac{f^\prime(\mathtt{x})}{
f(\mathtt{x}) - \log \mu_d}\,d\mathtt{x}
\leq \int_{\log z}^{\log a} \frac{d\mathtt{x}}{\mathtt{x}-
\log \frac{\mu_d}{\mu_{d-1}}}.\]
Integrate both sides and collect terms. 
\qed

This gives the following asymptotic bounds for $\kappa(a)$.

\begin{corr}
For every $\delta>0$ there exists an $a_\delta$ such
that for all $a> a_\delta$,
\[ \log \mu_{d-1} + \log a \leq \kappa(a) \leq
  \log \mu_{d-1} + (1+\delta) \log a .\]
\label{corrA}  %%ZXZ[corrA]
\end{corr}
\Pr
Substitute the upper bound $\kappa(z) \leq \log \mu_d + \log z$
in the right hand side of Theorem \ref{thmA} to find that
for $a>z>\sfrac{\mu_d}{\mu_{d-1}}$,
\begin{equation}
\kappa(a) - \log \mu_d \leq
\left(\log a -\log\frac{\mu_d}{\mu_{d-1}} \right) 
\left( \frac{\log z}{\log z-\log\frac{\mu_d}{\mu_{d-1}}}  \right) .
\label{eqn39}
\end{equation}
Let $\delta>0$ and choose $z$ so large that 
\[ 1 \leq \frac{\log z}{\log z-\log\frac{\mu_d}{\mu_{d-1}}} \leq
1 + \delta . \]
Put $a_\delta=z$ so that $a>a_\delta$.
Then for $a>a_\delta$ equation (\ref{eqn39}) becomes
\[ \log a + \log \mu_{d-1} \leq \kappa(a) \leq
\log \mu_d + (1+\delta) 
\left(\log a -\log\frac{\mu_d}{\mu_{d-1}} \right), \]
where we recall the lower bound from equation \Ref{eqn3.1}.
Expand and simplify the right hand side to obtain
\[ \log a + \log \mu_{d-1} \leq \kappa(a) \leq
 (1+\delta)\log a + (1+\delta)\log \mu_{d-1} 
- \delta\log \mu_d . \]
This may be put in the form
\[ \log a + \log \mu_{d-1} \leq \kappa(a) \leq
\log\mu_{d-1} + (1+\delta)\log a 
+ \delta \log \frac{\mu_{d-1}}{\mu_d}. \]
Since $\frac{\mu_d}{\mu_{d-1}} > 1$ the last
term may be discarded.
\qed

That is, $\kappa(a)$ is asymptotic to $\log\mu_{d-1} + \log a$
in the sense of Corollary \ref{corrA}.

Since $\kappa (a) = \log \mu_d$ if $a \leq 1$ 
and $\kappa (a) > \log \mu_d$
if $a>a_c^o$, there is an \emph{adsorption transition}
in the model where ${\cal E}(a)$ becomes strictly positive.  
These results are taken together in the following theorem.

\begin{theo}
There is a critical point $a_c^o$ 
separating a phase of free positive walks from a phase
of adsorbed walks. 

For $a < a_c^o$, $\kappa(a)= \log \mu_d$.

If $a > a_c^o$, then 
\[ \max\{\log \mu_d , \log \mu_{d-1} + \log a\} 
\leq \kappa(a) 
\leq \max\{\log \mu_d, \log \mu_d + \log a \} . \]
Moreover, $\kappa(a)$ is asymptotic to $\log\mu_{d-1} + \log a$ 
for large $a$, and 
$a_c^o \in \left[ 1,\sfrac{\mu_d}{\mu_{d-1}} \right]$. \qed
\label{thmAAa}  %%ZXZ[thmAAa]
\end{theo}

\subsection{Bounds on $\lambda (y)$}

We first recall some earlier results for pulled self-avoiding
walks (see section 2.3 in reference \cite{Rensburg2009}). 

It is known that $\lambda (y)$ is a non-analytic 
function of $y$ with a singularity at $y=y_c^o$ where 
$1 \leq y_c^o \leq \mu_d$.  It is also known that $\lambda(y)
= \log \mu_d$ if $y \leq 1$.  Bounds for $y>1$ are also known, 
and are given by
\begin{equation}
\max\{\log \mu_d , \log y\} \leq \lambda (y)
\leq \log \mu_d + \log y .
\label{eqn3.10}   %%ZXZ[eqn3.10]
\end{equation}

The number of tails from the origin of length $n$ and height
of last vertex $h$ is $c_n^+(0,h)$.  The number of unfolded
tails of length $n$ is $c_n^\ddagger (0,h)$.

Then the partition function of unfolded tails is given by
\begin{equation}
T_n^\ddagger(y) = \sum_h c_n^\ddagger(0,h)\,y^h .
\end{equation}

The average height of a pulled tail is given by
\begin{equation}
\langle h \rangle_n = y\sfrac{d}{dy}
\log \sum_h c_n^\ddagger(0,h)\,y^h .
\end{equation}
Since $\sfrac{1}{n} \log T_n^\ddagger (y)$ is a sequence
of convex functions converging to a convex limit
$\lambda(y)$, one may interchange the limit and the
derivative almost everywhere in
\begin{equation}
{\cal E}^\lambda (y) = \lim_{n\to\infty} \sfrac{1}{n}
\langle h \rangle_n = \lim_{n\to\infty} \sfrac{y}{n}
\sfrac{d}{dy} \log T_n^\ddagger (y) = y \sfrac{d}{dy}
\lambda(y) .
\end{equation}
Similarly to equation \Ref{eqn3.3}, one may define
left- and right-densities of the height of the endpoint:
\begin{equation}
{\cal E}^\lambda_-(y) = y\sfrac{d^-}{dy} \lambda(y),\qq
{\cal E}^\lambda_+(y) = y\sfrac{d^+}{dy} \lambda(y).
\end{equation}
These exist for every $y>0$, while ${\cal E}^\lambda_-(y)$
is lower semi-continuous and ${\cal E}^\lambda_+(y)$
is upper semi-continuous.  Notice that
${\cal E}^\lambda_- (y) \leq {\cal E}^\lambda_+(y)$
generally and ${\cal E}^\lambda_- (y) 
= {\cal E}^\lambda_+(y) = {\cal E}^\lambda(y)$ for 
almost all $y>0$.

The Legendre transform of $\lambda(y)$ is defined by
\begin{equation}
\log \C{P}^\lambda(\beta) = \inf_{\beta>0}
\LC \lambda(y) - \beta \log y \RC.
\end{equation}
The function $\log \C{P}^\lambda(\beta)$ is a concave
function of $\beta$ on $[0,1]$ (see for example
reference \cite{Rensburg2000B}).  The inverse transform
is given by
\begin{equation}
\lambda(y) = \sup_{\beta\in[0,1]} \LC
 \log \C{P}^\lambda(\beta) +\beta \log y \RC .
\label{eqn4-7}  
\end{equation}
Since $\lambda(y)$ is defined for every $y\geq 0$,
there exists a $\beta^*\in[0,1]$ solving equation
\Ref{eqn4-7}.

Next, define the function $h_n^* = \lfl \langle h \rangle_n\rfl$.
Arguments similar to those leading to Theorem \ref{thm3-7} then give
the following theorem:  

\begin{theo}
For almost every value of $y>0$,
$$\lim_{n\to\infty} \sfrac{1}{n}\, h^*_n 
= {\cal E}^\lambda(y) = \beta^*$$
where $\beta^*$ realizes the supremum in equation
\Ref{eqn4-7}. \qed
\label{thm4-7}
\end{theo}

A lower bound on $T_n(y)$ may be obtained by 
only considering unfolded tails
of length $n-h^*+1$ which end in a vertex at height $1$,
followed by a sequence of $h^*-1$ vertical edges to end
at height $h^*$.  This shows that
$c_{n-h^*_n+1}^t(1) \, y^{h_n^*} \leq T_n(y)$.
Take logarithms, divide by $n$ and let $n\to\infty$ to obtain
\begin{equation}
\L 1 - \C{E}^\lambda (y) \R\, \log \mu_d
 + \C{E}^\lambda(y)\, \log y \leq \lambda (y) ,\qq
\hbox{for almost every $y \geq y_c^o$},
\end{equation}
and where $\C{E}^\lambda (y) = y\frac{d}{dy} \lambda(y)$.

This gives the following lemma:

\begin{lemm}
Suppose that $f(x) = \lambda(e^x)$.  Then for almost every
$x > \log y_c^o$,
\[ f(x) - \log \mu_d 
\geq \frac{d}{dx} f(x)\, \left(x - \log\mu_d \right) . \]
\label{lemAy}  %%ZXZ[lemAy]
\end{lemm}

Integrate the inequality in lemma \ref{lemAy} from
$\log z$ to $\log y$ where $z>\mu_d$.
This gives the following theorem.

\begin{theo}
The free energy of pulled walks satisfies the
following inequality:
\[ \frac{\lambda(y) - \log \mu_d}{\log y -\log\mu_d}
\leq \frac{\lambda(z) - \log \mu_d}{\log z-\log\mu_d}\]
whenever $y > z> \mu_d$.
\label{thmAy}  %%ZXZ[thmAy]
\end{theo}
\Pr
Separate variables in Lemma \ref{lemAy} and integrate 
over $(\log z,\log y)$ where $z>\mu_d$.  This gives
\[ \int_{\log z}^{\log y} \frac{f^\prime(\mathtt{x})}{
f(\mathtt{x}) - \log \mu_d}\,d\mathtt{x} \leq 
\int_{\log z}^{\log y} \frac{d\mathtt{x}}{\mathtt{x}- \log\mu_d }.\]
Integrate both sides and collect terms. 
\qed

Use the upper bound $\lambda (y) \leq \log \mu_d + \log y$
for $y \geq 1$ in 
Theorem \ref{thmAy}.  This shows that
\begin{equation}
\lambda(y) - \log \mu_d \leq
\left(\log y -\log \mu_d\right) 
\left( \frac{\log z}{\log z-\log \mu_d}  \right) .
\label{eqn39Y}
\end{equation}
Let $\delta>0$ and choose $z\geq 1$ so large that 
\[ 1 \leq \frac{\log z}{\log z-\log \mu_d} \leq
1 + \delta . \]
Put $y_\delta = z$.
Since $\lambda (y) \geq \log y$, the above bound in
equation (\ref{eqn39Y}) becomes
\[ \log y \leq \lambda(y) \leq
\log \mu_d + (1+\delta) 
\left(\log y -\log \mu_d \right) \]
whenever $y > y_\delta$. Expand and simplify the right hand 
side to obtain
\[ \log y \leq \lambda(y) \leq
(1+\delta)\log y - \delta \log \mu_d. \]
This is true for all $y > z$.  Hence, given $\delta>0$
there is an $y_\delta$ such that these bounds are true
for all $y > y_\delta$.  Since $\mu_d>1$ the last term
in the upper bound is negative and may be discarded.
This gives

\begin{corr}
For every $\delta>0$ there exists an $y_\delta$ such
that for all $y> y_\delta$,
\[ \log y \leq \lambda(y) \leq
  (1+\delta) \log y  .\]
\label{corrAy}  %%ZXZ[corrAy]
\qed
\end{corr}

Thus, there is a phase transition where walks become
ballistic.  This follows in particular because
$\lambda(y) = \log \mu_d$ if $y \leq 1$ and from
the bounds in equation (\ref{eqn3.10}).  These results
may be taken together in the following theorem.

\begin{theo}
There is a critical point $y_c^o$ separating a free phase 
from a ballistic phase in the limiting free energy $\lambda(y)$ 
of (vertically) pulled positive walks.  

For $y < y_c^o$, $\lambda (y) = \log \mu_d$.

If $y > y_c^o$, then 
\[ \max\{\log \mu_d , \log y\} \leq \lambda (y)
\leq  \log \mu_d + \log y  . \]
Moreover, $\lambda(y)$ is asymptotic to $\log y$ for large
$y$, and $y_c^o \in \left[ 1,\mu_d \right]$. \qed
\label{thmAAy}  %%ZXZ[thmAAy]
\end{theo}

By introducing the reduced pulling force $\hat{f}$ in the model by setting
$y=\exp \hat{f}$, the last theorem states that there is a critical reduced force
$\hat{f}_c \in \left[ 0,\log \mu_d \right]$ such that 
for $\hat{f}>\hat{f}_c$ the model is ballistic.  Numerical simulations
suggest that $\hat{f}_c=0$ \cite{Rensburg2009}.  See also \cite{IoffeVelenik}.

\section{Phase transitions and the phase diagram of pulled
adsorbing walks}
\label{sec:phasediag}

In this section the phase diagram of a pulled adsorbing walk
is examined.  The free energy of the model is $\kappa(a,y)$
(see equation \Ref{FE}) and $\kappa(a,y) = \max\LC \kappa(a),
\lambda(y) \RC$.  By Theorem \ref{convexity} $\kappa(a,y)$ 
is a convex function of $\log a$ and $\log y$.

We next examine the properties of $\kappa(a,y)$ by
using the results of section 3 on $\kappa(a)$ and $\lambda(y)$.

If $a < a_c^o$ and $y < y_c^o$ the $\kappa(a) = 
\lambda (y) = \log \mu_d$.  Thus, $\kappa(a,y) = 
\log \mu_d$ in this regime, which is a \emph{free} phase.

If $a<a_c^o$ is fixed, and $y< y_c^o$ is increased
then there is a critical point at $y_c^o$ where
$\lambda(y)> \log \mu_d$ so that the model transitions
into a \textit{ballistic} phase (where $\kappa(a,y)$ is only 
dependent on $y$).  This shows that the line segment
$0 \leq a < a_c^0$ and $y=y_c^o$ is a critical line
in the phase diagram, separating a free phase from a 
ballistic phase.

Similarly, if $y<y_c^o$ and $a<a_c^o$ is increased
then there is a critical point at $a_c^o$ where 
$\kappa(a) > \log \mu_d$ so that the model goes 
through a transition to an \textit{adsorbed} phase.  This shows
that the line segment $a \leq y < y_c^o$ and $a=a_c^o$
is a critical line in the phase diagram, separating a 
free phase from an adsorbed phase. 

Next, consider the cases that either $a>a_c^o$ or
$y>y_c^o$.

\begin{lemm}
For $a > a_c^o$ there exists a real number $y_c(a)$, 
$1 \le y_c(a) \le e^{\kappa (a,1)}$,
such that the free energy is equal to $\kappa (a,1)$  for 
$y < y_c(a)$ and, for $y > y_c(a)$,
$$\kappa (a,y) \ge \max [\kappa (a,1), \log y].$$
\label{lemma5}
\end{lemm}
\Pr
Fix $a > a_c^o$ and note that $\kappa(a,1)=\kappa(a)$.  
For $y \le 1$, $\kappa(a,y) \le \kappa(a,1)$
by monotonicity.  Also 
$$\kappa(a,y) \ge \lim_{n\to\infty} \sfrac{1}{n} \log L_n(a) 
= \kappa (a) = \kappa(a,1)$$
and hence $ \kappa (a,y) = \kappa (a,1)$ for $y \le 1$. 

For $y \ge 1$, $\kappa(a,y) \ge \kappa(a,1)$ by monotonicity.  By extracting the single term in the partition function, where the 
walk is a straight line normal to the surface,
$\kappa(a,y) \ge \log y$.  Hence $\kappa(a,y)$ is singular 
at a point $y = y_c(a)$ where
$1 \le y_c(a) \le e^{\kappa(a,1)} = e^{\kappa (a)}$.  
\qed

The scaled average height of the last vertex, 
$\sfrac{1}{n} \langle h \rangle_n 
 = \langle z_n \rangle$, is given by 
\begin{equation}
\frac{\langle h \rangle}{n} 
= \frac{\sum_{v,h} h\, c_n(v,h)\, a^v y^h} {
\sum_{v,h} n\, c_n(v,h)\, a^vy^h} 
= \sfrac{1}{n} \sfrac{\partial}{\partial \log y}
\log \sum_{v,h} c_n(v,h)\, a^v y^h.
\end{equation}
In the $n\to\infty$ limit this becomes
\begin{equation}
\widehat{h} = \lim_{n\to\infty} \frac{\langle h \rangle }{n} 
= \frac{\partial \kappa(a,y)}{\partial \log y},\qq
\hbox{for almost every $(a,y)$},
\end{equation}
where we have used the convexity of $\kappa (a,y)$ when 
we interchanged the order of the limit and the derivative.

Since $\kappa(a,y) = \max\LC \kappa(a),\lambda(y) \RC$
it follows from Lemma \ref{lemma5} that if $a>a_c^o$
and $y < y_c(a)$ then the free energy is 
independent of $y$ and is equal to $\kappa(a)$.  The 
limiting scaled average height of the last vertex, 
$\widehat{h} =  \lim_{n\to \infty} \sfrac{1}{n}
\langle h \rangle_n =0$ and $\langle h \rangle_n = o(n)$
in this regime.   

For $a > a_c^o$ and $y > y_c(a)$ the free energy is given 
by $\kappa(a,y) = \lambda(y)$ and this is strictly increasing
with $y$ so that $\widehat{h} > 0$.  That is, the walk is in 
a ballistic phase.

These results, together with Theorem \ref{theo:thermolim}, 
show  that we have phases where the walk is adsorbed and where 
the walk is ballistic, but there is no phase in which the 
walk is partly adsorbed and partly ballistic (such phases
are present in a directed model of an adsorbing copolymer
pulled in the middle \cite{IJvR2012}).  

We can obtain useful upper bounds on $\kappa(a,y)$ 
by noting that in the adsorbed phase the free energy is 
independent of $y$ and in the desorbed phase it is 
independent of $a$.  Therefore, in the adsorbed phase 
\begin{equation}
\kappa (a,y) \le \log \mu_d + \log a
\label{eqn:bounda}
\end{equation}
and in the desorbed phase 
\begin{equation}
\kappa (a,y) \le \log \mu_d + \log y.
\label{eqn:boundy}
\end{equation}

\begin{lemm}
For $a > a_c^0$ the phase boundary $y=y_c(a)$ between the adsorbed phase and the desorbed ballistic phase
satisfies the bounds
$$ \max \left[ 1,a\, \sfrac{\mu_{d-1}}{\mu_{d}} \right] \le y_c(a) \le a\, \mu_d.$$
\end{lemm}
\Pr
Fix $a > a_c^o$ and suppose that $y \ge 1$.  
If $y \le y_c(a)$ we are in the adsorbed phase and 
$\kappa(a,y) \le \log \mu_d + \log a$.  But
$\kappa(a,y) \ge \log y$.  Hence if $ \log y 
> \log \mu_d + \log a$ we have a contradiction and 
we are in the ballistic phase.  Therefore 
$y_c(a) \le a\, \mu_d$.  

Similarly if $y \ge y_c(a)$ then $\kappa(a,y) 
\le \log \mu_d + \log y$.  But $\kappa(a,y) 
\ge \max[\log \mu_d, \log \mu_{d-1} + \log a]$.  
Hence $y_c(a) \ge \max \left[ 1,a\ \sfrac{\mu_{d-1}}{\mu_d} \right]$.
\qed

\begin{figure}[h!]
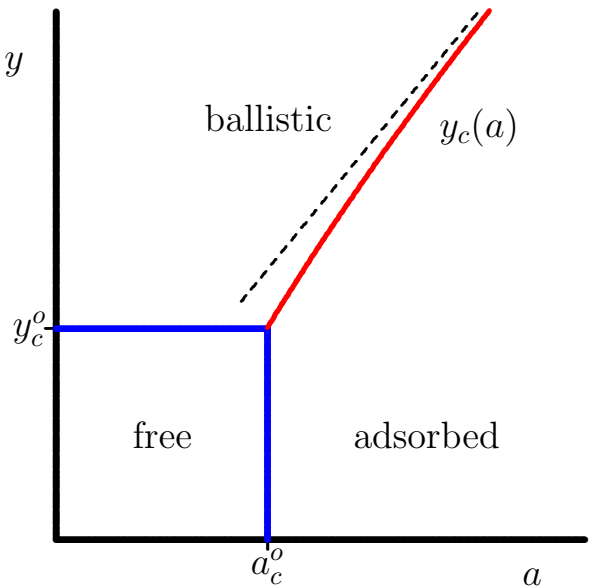

\beginpicture
\setcoordinatesystem units <2pt,2pt>
\setplotarea x from -40 to 100, y from -10 to 100

\color{black}
\setplotsymbol ({$\cdot$})
\plot -2 40 0 40 / \plot 40 -2 40 0 /

\setdashes <3pt>
%\plot 45 35 100 80 /
\plot 35 45 80 100 /
\setsolid

\setplotsymbol ({\tiny$\bullet$})
\plot 0 100 0 0 100 0 /

\color{black}
\put {\Large$y$} at -8 90
\put {\Large$a$} at 90 -7
\put {\Large$y_c^o$} at -5 40
\put {\Large$a_c^o$} at 40 -5
\put {\Large$y_c(a)$} at 80 78

\put {\Large\hbox{free}} at 20 20 
\put {\Large\hbox{ballistic}} at 40 80 
\put {\Large\hbox{adsorbed}} at 70 20

\color{blue}
\plot 40 0 40 40 0 40 /

\setplotsymbol ({\LARGE$\cdot$})
\color{red}
\setquadratic
\plot 40 40 59 69 82 100 /
\setlinear

\color{black}
\normalcolor
\endpicture
\caption{The phase diagram of pulled adsorbing walks.
The critical curve $y_c(a)$ separating the ballistic
and adsorbed phases is asymptotic to the dashed
line in the sense that $\log y_c (a) = 
\log (a\mu_{d-1}) + o(\log(a))$.  There are numerical
results indicating that $y_c^o=1$ (see also
\cite{IoffeVelenik}).  It is known
that $a_c^o>1$ \cite{HTW1982}.}
\label{figure6}
\end{figure}

Taking the results above together gives the phase diagram
in Figure \ref{figure6}.

%%%%%%%%%%%%%%%%%%%%%%%%%%%%%%%%%%%%%%%%%%%%%%%%%%
%%%%%%%%%%%%%%%%%%%%%%%%%%%%%%%%%%%%%%%%%%%%%%%%%%

The phase boundary $y_c(a)$ is monotone strictly increasing, as
shown below.

\begin{lemm}
The phase boundary $y=y_c(a)$ is monotone strictly increasing.
\label{monotone}
\end{lemm}
\Pr
Take two points, $(a_1,y_1)$ and $(a_2,y_2)$ on the phase boundary.  
Suppose that $a_2 > a_1$.  Then $\kappa(a_1)=\lambda(y_1)$ 
and $\kappa(a_2)=\lambda(y_2)$.  But $\kappa(a)$ is 
strictly increasing so $\kappa(a_2) > \kappa(a_1)$ and 
therefore $\lambda(y_2)>\lambda(y_1)$.  But $\lambda(y)$ 
is also strictly increasing so $y_2 > y_1$. \qed

Finally, it is a corollary of Corollaries \ref{corrA}
and \ref{corrAy} that $y_c(a)$ is asymptotic to
$a\,\mu_{d-1}$ as $a\to\infty$.

To see this, notice that the critical curve $y_c(a)$ is
given by the solution of $\lambda(y) = \kappa(a)$, by
definition of $\kappa(a,y)$. That is $\lambda(y_c(a))
= \kappa(a)$ for almost all $a>a_c^o$ by monotonicity of
$y_c(a)$.

Choose $\delta>0$.  Since $y_c(a)$ strictly increases
with $a$, there is an $a_1$ such that if $a>a_1$, then
$y_c(a) > y_\delta$ in Corollary \ref{corrAy}.  Increase
$a$, if necessary, until $a>\max\{a_1,a_\delta\}$ in Corollary
\ref{corrA}.  Then both the conditions of Corollaries
\ref{corrA} and \ref{corrAy} are valid, and so 
\begin{eqnarray}
&\log \mu_{d-1} + \log a 
\leq \kappa(a)
\leq \log \mu_{d-1} + (1+\delta)\log a, \nonumber \\
&\log y_c(a) 
\leq \lambda(y_c(a)) 
\leq (1+\delta) \log y_c(a).
\label{eqnarray1}
\end{eqnarray}
Noting that $\kappa(a) = \lambda(y_c(a))$ for almost 
all $a>a_c^o$, it follows that
\begin{equation}
\log \mu_{d-1} + \log a \leq (1+\delta) \log y_c(a).
\end{equation}
Rearrange this into $\sfrac{\log y_c(a)}{\log (a\mu_{d-1})}
\geq \sfrac{1}{1+\delta}$ and take $a\to\infty$.  Then
$y_c(a) \to \infty$ by Lemma \ref{monotone} so that
for any small $\delta>0$
\begin{equation}
\liminf_{a\to\infty} \frac{\log y_c(a)}{\log (a\mu_{d-1})} 
\geq \frac{1}{1+\delta} .
\label{eqnliminf1}
\end{equation}

On the other hand, it similarly follows from equation
\Ref{eqnarray1} that 
\begin{equation}
\log y_c(a) \leq \log \mu_{d-1} + (1+\delta)\log a.
\end{equation}
Rearrange this into $\sfrac{\log y_c(a)}{\log (a\mu_{d-1})}
\leq 1 + \sfrac{\delta \log a}{\log a + \log \mu_{d-1}}$ 
and take $a\to\infty$. Then
$y_c(a) \to \infty$ by Lemma \ref{monotone} so that
for any small $\delta>0$
\begin{equation}
\limsup_{a\to\infty} \frac{\log y_c(a)}{\log (a\mu_{d-1})} 
\leq 1 + \delta.
\label{eqnlimsup1}
\end{equation}

Taking $\delta \to 0^+$ in equations \Ref{eqnliminf1}
and \Ref{eqnlimsup1} gives the following theorem.

\begin{theo}
The critical curve $\log y_c(a)$ is asymptotic to 
$\log (a\mu_{d-1})$.  That is,
\[ \lim_{a\to\infty} \frac{\log y_c(a)}{\log (a\mu_{d-1})} 
= 1 . \]
Hence $\log y_c(a) = \log (a\mu_{d-1}) + 
o(\log a)$ as $a\to\infty$. \qed
\end{theo}

\section{Discussion}
\label{sec:discuss} \setcounter{equation}{0}

We have considered a self-avoiding walk model of polymer adsorption 
where the adsorbed walk is subject to a force, applied normal to the surface, 
that tends to desorb the walk.  We have proved several rigorous results 
about this model, including the existence of the limiting free
energy, the convexity of the free energy and connections between this two
variable model and the two one variable models where (a) the walk adsorbs but 
without an applied force, and (b) the walk is subject to a force but does not adsorb.

We have shown that there is a phase boundary in a relevant two variable 
space, proved some qualitative results about the boundary and derived 
bounds on its location. Our results show that there are only three phases
in the model, namely a free phase, a ballistic phase and an adsorbed
phase.  There is no phase which exhibits both the characteristic of
being adsorbed and being ballistic in this model, in contrast with
results on a directed path model (with the pulling force in the middle)
\cite{Rensburg2010,Rensburg2010A}.

There are interesting questions about reentrant phase boundaries in models such as 
the one considered here \cite{Krawczyk2005}.  The asymptotic result about the 
shape of the phase boundary in the large $a$ limit,
derived in the last section,  in fact establishes reentrance
in this model in three and more dimensions. 
By corollary \ref{corrA} and lemma \ref{monotone} 
for any fixed $\delta > 0$  we can take $a$ large enough that
\begin{equation}
\fl
\hspace{1cm}
\log \mu_{d-1} + \log a \leq (1+\delta) \log y_c(a) \q\hbox{and}\;
\log y_c(a) \leq \log \mu_{d-1} + (1+\delta ) \log a.
\end{equation}
If we connect to physical variables by writing $a = \exp [-\epsilon /kT]$
and $y = \exp [f/kT]$ where $\epsilon < 0$ is the energy associated with a visit, $f$ is the 
applied force, $k$ is Boltzmann's constant and $T$ is the temperature, this relation
becomes
\begin{equation}
(1-\delta)\,\L kT \log \mu_{d-1} - \epsilon \R \le f 
\le kT \log \mu_{d-1} - (1+ \delta) \, \epsilon
\end{equation}
for sufficiently small values of $T$.  If $d\geq 3$ then $\mu_{d-1}>1$.  This
shows that the critical force is an increasing function of $T$ at sufficiently low $T$
provided that $d\geq 3$. This establishes reentrance.  Moreover,
\begin{equation}
{\cal S}_d = \lim_{T\to 0} \frac{df}{dT} = k \log \mu_{d-1}
\end{equation}
which is the ground state entropy, and this is strictly positive if $d\geq 3$
(but ${\cal S}_2 = 0$).

This model has several interesting extensions.  What happens if the force
is applied at an angle to the surface?  What happens if the walk or the surface
has (regular) heterogeneity?  We intend to explore these questions in subsequent papers.

\section*{Acknowledgements}
This research was partially supported by NSERC of Canada.

\end{document}